\documentclass{article}

\PassOptionsToPackage{numbers, compress}{natbib}

\usepackage[final,sglblindworkshop]{neurips_2026}
\workshoptitle{LIGHT - Lightweight Trustworthy Foundation Models}

\usepackage[utf8]{inputenc}
\usepackage[T1]{fontenc}
\usepackage{hyperref}
\usepackage{url}
\usepackage{booktabs}
\usepackage{amsmath}
\usepackage{amsfonts}
\usepackage{nicefrac}
\usepackage{microtype}
\usepackage{xcolor}
\usepackage{colortbl}
\usepackage{graphicx}
\usepackage{multirow}
\usepackage{float}

\definecolor{goodblue}{RGB}{80,140,215}
\definecolor{badred}{RGB}{225,90,80}

\title{From Clicks to Intent: Distilling LLM Intent Understanding into Compact, Deployable Models in Financial Services}
\author{%
  Dianjing Fan\thanks{Equal contribution.} \\
  Capital One \\
  McLean, VA, USA \\
  \texttt{dianjing.fan@capitalone.com} \\
  \And
  Yao Li\footnotemark[1] \\
  Capital One \\
  New York, NY, USA \\
  \texttt{yao.li@capitalone.com} \\
  \And
  Kyaw Hpone Myint\thanks{Corresponding author.} \\
  Capital One \\
  Boston, MA, USA \\
  \texttt{kyaw.hponemyint@capitalone.com} \\
  \AND
  Dwipam Katariya \\
  Capital One \\
  McLean, VA, USA \\
  \texttt{dwipam.katariya@capitalone.com} \\
  \And
  Alexandre Day \\
  Capital One \\
  McLean, VA, USA \\
  \texttt{alexandre.day@capitalone.com} \\
  \And
  Pranab Mohanty \\
  Capital One \\
  McLean, VA, USA \\
  \texttt{pranab.mohanty@capitalone.com} \\
  \And
  Giri Iyengar \\
  Capital One \\
  McLean, VA, USA \\
  \texttt{giridharan.iyengar@capitalone.com} \\
}

\begin{document}

\maketitle

\begin{abstract}
Large Language Models (LLMs) can extract meaning from noisy, unstructured data that traditional methods struggled to interpret, including the raw user clickstream data that industrial platforms collect at scale. Such models can infer user intent from behavioral logs, but their size, latency, and inconsistency make them impractical to deploy directly in high-volume, regulated settings. In addition, existing production systems for capturing user intent are often ad hoc and narrow, lacking the flexibility to support both quantitative downstream recommendations and qualitative understanding at scale. In this work, we present a system that distills the intent-understanding capability of an LLM into lightweight, deployable models suited to these constraints and demonstrate its applicability for personalization in financial services. A self-supervised Transformer encodes raw web clickstreams into a 64-dimensional session level embedding. An LLM then builds an interpretable intent taxonomy and labels each session. These labels are distilled into a lightweight classifier that runs directly on the embedding, with no LLM calls or text encoding at inference. The model produces two outputs: the dense embedding drives quantitative prediction, while the distilled labels add human-readable intent for explainability and governance. In production, the distilled pipeline runs orders of magnitude faster than direct LLM inference, cutting the compute cost of large-scale annotation with only a 7\% drop in labeling quality, and it scales to millions of sessions per day. On downstream tasks, the session embedding improves macro Recall@1 by 1.88\% and cuts Log Loss by 13.38\% over baselines on the mobile homepage tile ranking task, and it outperforms the LLM's labels by 4.3\% micro F1 on the user conversion prediction task. This results in two compact models that run at production scale while keeping the accuracy and interpretability that a regulated financial services deployment requires.
\end{abstract}

\section{Introduction}

Industrial platforms serve their users through digital channels and record massive volumes of clickstream data such as timestamps, page views, and user actions, but these logs capture \emph{what} users did, not \emph{why}: the underlying intent stays latent. Large Language Models (LLMs) can move beyond the surface-level interactions to infer the user intent directly from the raw logs, an understanding that drives personalization, segmentation, and analytics~\citep{Li2019MIND, Chen2022ICLRec, Kwon2024TaxonomyGuidedRec}.
Deploying LLMs directly at production scale, however, is impractical: they are large, slow, and costly to run over millions of sessions, and prompting them directly returns inconsistent, free-form labels that cannot be explained or audited to the standard that a regulated industry requires. The open question is how to preserve an LLM's intent-understanding capability while meeting the compute, latency, and governance constraints in real-world deployment.

We study this question in financial services, a setting that stresses all the constraints at once: traffic is high-volume, decisions are regulated, and the behavioral signals are underutilized. In particular, intent from pre-login web browsing carries significant weight: most financial product exploration, comparison, and even application initiation occur on the public website before authentication, making pre-login browsing one of the most valuable signals available. Current pre-login intent features are ad hoc and narrow (e.g., logging that ``a user visited credit card pages''), failing to capture more nuanced behavioral signals, while running an LLM per session to understand each intent is infeasible at scale. Neither serves both the quantitative tasks (ranking, recommendation) and the qualitative use cases (segmentation, analytics, audit) that a regulated financial industry requires.

We address this with a system that distills an LLM's intent-understanding capability into two compact models that run in production at scale, and we demonstrate them for personalization in financial services. To model the pre-login signal while evaluating post-login outcomes, we link the two through first-party cookies\footnote{All clickstream data used in this study is collected from a single first-party domain and is handled under internal data governance policies.} and shared login credentials. Our contributions are:

\begin{itemize}
  \item \textbf{A compact, dual-output intent model.}
  A self-supervised Transformer encodes multi-modal web clickstream events into a single session level embedding, and an LLM-generated intent taxonomy is distilled into a lightweight classifier over that same embedding. One shared representation yields two outputs: a dense embedding for quantitative tasks and human-readable labels for qualitative understanding.
  \item \textbf{Clustering-enhanced taxonomy generation.}
  A clustering layer over session embeddings provides stratified sampling that improves iterative LLM taxonomy generation~\citep{Wan2024TnTLLM}, so the LLM sees the full diversity of user behavior rather than a random subset. The taxonomy also constrains the LLM to a fixed label set instead of free-form text, giving stable, consistent intent labels that a regulated setting requires.
  \item \textbf{Embedding-space distillation for LLM-free inference.}
  Distilling the LLM's labels into a classifier that operates on the fixed-dimensional embedding, rather than on raw text sequences, removes LLM calls and text encoding at inference and scales to millions of daily sessions.
\end{itemize}

\section{Related Work}

\subsection{User Behavior Representation Learning}
\label{sec:clickstream-repr}

Modeling user behavior sequences drives personalization in modern recommender systems, and methods differ primarily in how sequential dependencies are captured.
Early approaches apply traditional machine learning, such as manual rule mining~\citep{Kim2011AssocRules}, Markov Chain models~\citep{Bernhard2016Markov}, and unsupervised clustering~\citep{Wang2016ClickstreamClustering, Olmezogullari2020Embeddings}, but capture only static or short-range patterns.
Recurrent architectures such as GRU4Rec~\citep{Hidasi2016GRU4Rec} add gated sequential encoders, yet are non-parallelizable and lose sensitivity to long-range dependencies.

Transformer-based architectures address these limitations with self-attention over behavior sequences.
SASRec~\citep{Kang2018SASRec} applies unidirectional self-attention to capture short- and long-term dependencies, while BERT4Rec~\citep{Sun2019BERT4Rec} adds bidirectional attention with masked-item prediction.
Industry systems scale these to multi-feature settings such as BST~\citep{Chen2019BST}, TransAct~\citep{Xia2023TransAct}, TRACE~\citep{Dutta2024TRACE}, STEP~\citep{Melnychuk2024TabularTransformers}, and TIMeSynC~\citep{Katariya2024TIMeSynC}.
However, these systems fuse raw features without incorporating enriched semantic representations such as pretrained text embeddings from page content, and none provides an interpretability layer over the modeled sequences: the resulting representations cannot be inspected, summarized, or grounded in human-readable form.

\subsection{LLM Taxonomy Generation and Distillation}
\label{sec:llm-taxonomy}

In recommender systems, intent representations have structured candidate generation~\citep{Zhu2018TDM}, modeled multiple user interests~\citep{Li2019MIND}, and encoded latent intents in sequential models~\citep{Chen2022ICLRec}, but only implicitly; an explicit, human-readable taxonomy better supports behavioral analysis, user segmentation, and LLM-based ranking~\citep{Kwon2024TaxonomyGuidedRec}.

Pre-LLM taxonomy generation used hand-crafted patterns~\citep{Broder2002WebSearchTaxonomy}, hierarchical topic modeling~\citep{Blei2010hLDA}, and embedding-based clustering~\citep{Zhang2018TaxoGen}, but lacked scalability or nuanced semantics.
LLMs instead produce fluent, interpretable labels over document groups~\citep{Pham2024TopicGPT, Wang2023GoalEx, Zhang2023ClusterLLM}.
For large corpora, TnT-LLM~\citep{Wan2024TnTLLM} refines a taxonomy through iterative minibatches with human-in-the-loop validation~\citep{Yildirim2024IntentTaxonomy}, and SPRINT~\citep{Lee2026SPRINT} validates LLM-inferred intents against recommendation performance. Because serving an LLM per record is too slow and costly for production, these systems do not deploy the teacher directly: they distill its labels into lightweight student classifiers~\citep{Wan2024TnTLLM, Wettig2025WebOrganizer, Lee2026SPRINT}, consistent with broader evidence that an LLM teacher can train a compact student at a fraction of the inference cost~\citep{Wang2021GPT3CanHelp, Hsieh2023DistillStepByStep}.

In summary, prior work forces the trade-off this paper targets: sequential and LLM-based methods are expressive but large and opaque, whereas lightweight hand-crafted features are cheap but shallow. Compact models that distill an LLM's intent understanding into a representation that is both interpretable and deployable at scale over multi-modal clickstream sequences in financial services remain unexplored.

\section{Method}

\subsection{Overview}

Our system produces two complementary outputs from pre-login web clickstream data: (1) \textbf{a dense session embedding} for downstream tasks, and (2) \textbf{human-readable intent labels} for interpretability.
In the first stage (Section~\ref{sec:session-embedding}), raw clickstream events are encoded into session embeddings via a self-supervised Transformer.
In the second stage (Section~\ref{sec:taxonomy}), an LLM-driven pipeline constructs an intent taxonomy from representative sessions and assigns each session to multiple categories.
The resulting labels are distilled into a lightweight classifier that consumes the embeddings and produces interpretable intent labels for scalable inference.

\begin{figure}[t]
  \centering
  \includegraphics[width=\linewidth]{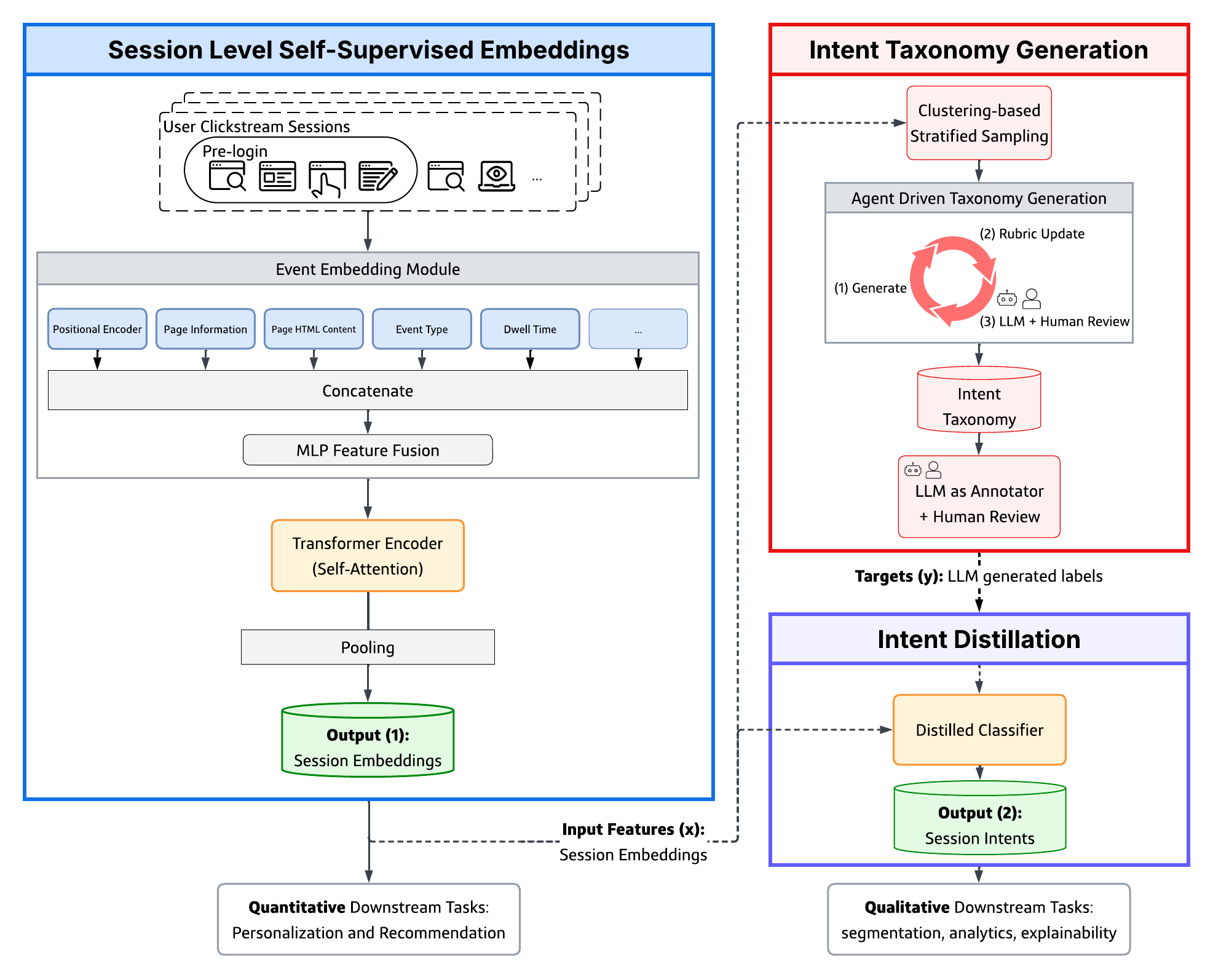}
  \caption{System architecture overview. \textbf{Left (Section~\ref{sec:session-embedding})}: multi-modal clickstream events are fused and encoded into session embeddings. \textbf{Right (Section~\ref{sec:taxonomy})}: clustering-based sampling, iterative taxonomy generation, label assignment, and distillation.}
  \label{fig:architecture}
\end{figure}

\subsection{Multi-Modal Session Embedding}
\label{sec:session-embedding}

The input to our model is pre-login web clickstream sessions: chronologically ordered sequences of events.
Let a session be $s = (x_1, \ldots, x_T)$ of length $T$ (the maximum session length, with shorter sessions padded), where each event $x_i$ carries $f$ heterogeneous features, including page information, dwell time, event type (click, scroll, view, etc.), and semantically enriched features such as page HTML content.
All features are encoded into a common $d$-dimensional space.

\textbf{Discrete features via learned embedding tables.}
Each discrete feature with vocabulary size $V$ is mapped to a $d$-dimensional vector by row lookup in a learned embedding matrix $E \in \mathbb{R}^{V \times d}$.

\textbf{Page content via a pretrained sentence encoder.}
To inject semantic content from the rendered page rather than the URL string alone, we pre-compute a $d_\text{url}$-dimensional vector for every URL by fetching its HTML, extracting the main text, and encoding it with a pretrained sentence transformer~\citep{Reimers2019SentenceBERT}, then project it to $d$ dimensions through a learned linear layer $W_\text{url} \in \mathbb{R}^{d \times d_\text{url}}$ with bias $b_\text{url}$.

\textbf{Sequence position via a learned positional embedding.}
A learned matrix $P \in \mathbb{R}^{T \times d}$ supplies one $d$-dimensional embedding per position, with row $P_i$ encoding position $i \in \{1, \ldots, T\}$.

\smallskip
\noindent Per event, the $f$ feature vectors are concatenated into an $fd$-dimensional vector and fused by an $\text{MLP}: \mathbb{R}^{fd} \to \mathbb{R}^d$; the positional embedding $P_i$ is then added, and a Transformer encoder $\mathcal{T}$ attends over the full sequence to produce contextualized event representations $(o_1, \ldots, o_T)$, with $o_i \in \mathbb{R}^d$.
Because the complete pre-login session is available at inference time rather than streamed incrementally, $\mathcal{T}$ is free to use either unidirectional or bidirectional self-attention, and the final session embedding is obtained by either last-hidden-state or mean pooling over $(o_1, \ldots, o_T)$.
The model is trained self-supervised: depending on the attention setting, it predicts the page title and URL of either the next event (causal) or masked events (bidirectional) from context.
This requires no labeled data and encourages the embedding to capture the semantic content and sequential structure of the session.
Depending on the model and downstream performance (see Section~\ref{sec:home}), the final model is a multi-layer Transformer encoder trained with bidirectional masked language modeling and mean pooling.

\subsection{Intent Taxonomy Generation and Distillation}
\label{sec:taxonomy}

The session embeddings capture user behavior densely but lack interpretability.
To bridge this gap, we construct an intent taxonomy using LLMs and distill it into a lightweight classifier that maps embeddings to human-readable labels, in three phases: clustering-based sampling, taxonomy generation, and label assignment and distillation.

\textbf{Clustering-based sampling.}
We cluster the session embeddings, comparing K-Means and HDBSCAN and tuning $K$ (or minimum cluster size) via silhouette score, Davies-Bouldin Index (DBI), and Calinski-Harabasz Index (CHI); we adopt K-Means and sample $N$ sessions per cluster, yielding $K \times N$ representative sessions that expose the LLM to the full diversity of user behavior rather than a skewed subset.
Clustering serves only as stratified sampling: clusters do not become taxonomy categories, since sessions far from centroids may carry multiple intents.
Each sampled session's raw clickstream is formatted as a sequence of \texttt{[dwell\_time] [event\_type] page\_title : page\_url} entries for LLM consumption.

\textbf{Taxonomy generation.}
Because page metadata changes frequently as content is produced dynamically, manually curating a taxonomy is impractical, motivating automated generation.
Inspired by TnT-LLM~\citep{Wan2024TnTLLM}, we construct a fixed intent taxonomy via an iterative minibatch approach with three stages---\textit{generate}, \textit{update}, and \textit{review}---which, unlike free-form labeling that yields inconsistently worded intents, constrains the label set and refines it as more data is observed.
The \textit{generation prompt} has the LLM produce an initial taxonomy of $I$ categories from the first minibatch, each with a name, description, and confidence score; the \textit{update prompt} then refines it over subsequent minibatches by adding, removing, merging, or splitting categories while maintaining the target count; and a final \textit{review prompt} checks completeness, mutual exclusivity, and clarity.
Repeating this generate--update--review cycle for $E$ epochs over the $K \times N$ sessions with minibatch size $B$ requires $\mathcal{O}(K \cdot N \cdot E / B)$ LLM calls.
Each prompt includes background context, task instructions, and formatting constraints (e.g., target category count, maximum label length); after each taxonomy, a manual human review refines the prompts so the result reflects domain expertise rather than LLM output alone.

\textbf{Label assignment and distillation.}
With the taxonomy fixed, and following findings that LLMs annotate well~\citep{Gilardi2023LLMAnnotator}, we use zero-shot prompting to classify each session into multiple taxonomy intents, each with a confidence score; micro intents (specific sub-goals) and chain-of-thought reasoning are also collected to support human review and prompt refinement.
For scalable inference, we distill these labels into a lightweight MLP student~\citep{Wan2024TnTLLM, Wettig2025WebOrganizer} that takes the session embedding as input and predicts each label's probability independently, trained with per-label binary cross-entropy on the LLM's confidence scores as soft targets to preserve the teacher's uncertainty and improve calibration.
At inference, the MLP produces intent labels at scale without separate text encoding or LLM calls.

\section{Experiments}
\label{sec:experiments}

\subsection{Experimental Setup}
\label{sec:experimental-setup}

\textbf{Web Clickstream Data.}
Our primary data source is internal clickstream data from our company's financial services website, within a single first-party domain with no cross-site tracking.
Each session is a chronologically ordered sequence of events (timestamps, page information, event type, cookies, device metadata), from the first page visit to idle timeout or logout; users may or may not log in.
We keep the pre-login portion, which captures intent before authentication.
The site receives approximately 10M sessions per day; filtering to sessions with both pre-login activity (e.g., product and servicing pages) and post-login activity (e.g., account management, payments, enrollment) yields approximately 100K per day.
To evaluate the embeddings and intents downstream, we join pre-login sessions with post-login signals (mobile homepage tile clicks and product conversions) via a temporal join, using four months of data with train, validation, and test splits partitioned by time to prevent leakage.

\begin{table}[H]
  \centering
  \caption{Example home screen tile categories.}
  \label{tab:tile-examples}
  \begin{tabular}{ll}
    \toprule
    Tile Category & Example \\
    \midrule
    Product & Internal product promotions \\
    Travel & Flight and hotel deals \\
    Shopping & External vendor shopping offers \\
    Referral & Referral rewards \\
    \bottomrule
  \end{tabular}
\end{table}

\textbf{Mobile Homepage Tile Ranking.}
The internal mobile home screen presents product and content tiles after login (example categories in Table~\ref{tab:tile-examples}), each powered by its own recommender and ordered by a global ranking model.
This validation dataset records which tiles each user taps, a direct signal of cross-platform post-login intent.
We frame it as multi-label classification: predict which tiles a user taps, i.e., $P(\text{click on tile } t \mid \mathbf{x})$ for each candidate tile $t$ given features $\mathbf{x}$.
The production model ranks tiles using post-login behavior (e.g., page-view sequences) and user context, with per-tile click-through rates as a simple baseline; none of these features capture pre-login signals.

\begin{table}[H]
  \centering
  \caption{Example user conversions.}
  \label{tab:conversion-examples}
  \begin{tabular}{l}
    \toprule
    Conversion \\
    \midrule
    Open credit card \\
    Apply for auto loan \\
    Enroll in paperless statements \\
    Book with travel portal \\
    \bottomrule
  \end{tabular}
\end{table}

\textbf{User Conversion.}
The other validation dataset tracks user conversions across mobile and web, i.e., which products or features users apply for or enroll in after pre-login browsing (examples in Table~\ref{tab:conversion-examples}).
We again frame it as multi-label classification, since a single session may lead to multiple conversions.
The task reflects a key financial-services constraint: for unauthenticated prospects with no prior relationship, pre-login clickstream activity is the \textit{only} available signal, as static user features and post-login behavior do not yet exist.

\subsection{Mobile Homepage Tile Ranking Results}
\label{sec:home}

We evaluate three session embedding variants, each added as a feature set on top of the baseline: (1) causal attention with last-hidden-state pooling, (2) causal attention with mean pooling, and (3) bidirectional attention with mean pooling.
The baseline is a LightGBM classifier trained one-vs-rest over tile categories, using click-through rate (CTR) and site retargeting features (SRF).
Both SRF and session embeddings derive from pre-login browsing: SRF aggregates page-visit counts by product category over rolling windows (9 dimensions), while the embeddings encode sequential patterns and content-level semantics.
We report macro Recall@K, the fraction of users whose clicked tile appears in the top $K$ predictions, and Log Loss, which measures calibration.
We use macro rather than micro metrics because of heavy class imbalance: the highest-frequency tile is a low-value servicing widget, while rarer tiles such as product offers and travel rewards drive greater revenue.

As shown in Table~\ref{tab:embedding-results}, all three variants improve recall over the baseline, confirming that pre-login session representations carry signal beyond coarse page-visit counts.
The bidirectional + mean pooling variant yields the largest gains, a 13.38\% Log Loss reduction and 1.88\% macro Recall@1 gain.
Bidirectional attention helps because the complete session is available at inference, so each event attends to both preceding and subsequent actions; mean pooling outperforms last-hidden-state pooling because the final event may be a low-signal action such as a redirect or bounce, and averaging avoids last-token bias.

\begin{table}[t]
  \centering
  \caption{Overall macro Recall@K and Log Loss on the home screen tile ranking task. To preserve business confidentiality, we report percentage changes relative to the baseline rather than absolute values; for Log Loss, negative values indicate improvement. Cell color encodes the change relative to the baseline (blue = improvement, red = degradation; darker is larger in magnitude, with the diverging scale saturating at a 6\% change).}
  \label{tab:embedding-results}
  \small
  \begin{tabular}{lcccc}
    \toprule
    Embedding Variant & R@1 & R@2 & R@3 & Log Loss \\
    \midrule
    CTR + SRF (baseline)   & ---     & ---     & ---     & ---     \\
    Causal + Last Hidden   & \cellcolor{goodblue!8}+0.56\% & \cellcolor{goodblue!8}+0.20\% & \cellcolor{goodblue!8}+0.70\% & \cellcolor{badred!14}+1.37\% \\
    Causal + Mean Pool.    & \cellcolor{goodblue!15}+1.50\% & \cellcolor{goodblue!12}+1.20\% & \cellcolor{goodblue!13}+1.32\% & \cellcolor{goodblue!60}$-$9.69\% \\
    Bidir. + Mean Pool.    & \cellcolor{goodblue!19}\textbf{+1.88\%} & \cellcolor{goodblue!12}\textbf{+1.16\%} & \cellcolor{goodblue!13}\textbf{+1.29\%} & \cellcolor{goodblue!60}\textbf{$-$13.38\%} \\
    \bottomrule
  \end{tabular}
\end{table}

\subsection{Intent Taxonomy and Conversion Prediction Results}
\label{sec:conversion-results}

Table~\ref{tab:taxonomy-examples} shows example intents at two granularity levels, $I{=}25$ and $I{=}35$.
The iterative process produces semantically coherent categories grounded in observed clickstream patterns rather than predefined rules. For clustering-based sampling, we select K-Means with $K{=}50$ based on silhouette score (0.535), Davies-Bouldin Index (1.478), and Calinski-Harabasz Index (19,250). Evaluating candidate sizes $I \in \{25, 35\}$, the annotator found $I{=}25$ too broad for actionable targeting; we selected $I{=}35$ for its balance between granularity and interpretability.

\begin{table}[t]
  \centering
  \caption{Example intents generated by the taxonomy pipeline at two granularity levels. $I{=}25$ serves as a compact catalog; $I{=}35$ adds finer-grained intents.}
  \label{tab:taxonomy-examples}
  \small
  \begin{tabular}{llc}
    \toprule
    Category & Intent & $I$ \\
    \midrule
    \multirow{2}{*}{Credit Cards}
      & Credit Card Preapproval & 25 \\
      & Credit Card Benefits & 35 \\
    \midrule
    \multirow{2}{*}{Bank Accounts}
      & Debit Card Information & 25 \\
      & Deposit Rate Comparison & 35 \\
    \midrule
    \multirow{2}{*}{Loans}
      & Auto Loan Prequalification & 25 \\
      & Vehicle Search & 35 \\
    \midrule
    \multirow{1}{*}{Business}
      & Business Credit Card Exploration & 25 \\
    \midrule
    \multirow{1}{*}{\shortstack[l]{Digital Service}}
      & Digital Banking Features & 25 \\
    \midrule
    \multirow{1}{*}{\shortstack[l]{General}}
      & Fraud Protection & 25 \\
    \bottomrule
  \end{tabular}
\end{table}

To validate LLM label quality, a domain expert manually reviews a stratified random sample spanning all intent categories, assessing whether the assigned labels capture the session's browsing intent and marking each as \textit{agree}, \textit{partially agree}, or \textit{disagree}.
The strict agreement rate is 97.0\% and the lenient rate (agree + partially agree) is 99.5\%, the single disagreement being a low-confidence edge case with no suitable label.
This confirms that LLM label assignment is reliable enough to serve as a teacher signal for distillation without exhaustive human annotation.

For the following experiments we report micro metrics: unlike the tile ranking task where rare clicks can drive high revenue and thus necessitate macro metrics, conversion frequency correlates directly with business impact, making frequency-weighted micro metrics appropriate for high-volume actions such as credit card applications.
The distilled student MLP reaches micro F1 of 0.822 against LLM-generated labels on held-out test data, showing that it reproduces the teacher's intent assignments while operating directly on the 64-dimensional embedding.

To validate the utility of our representations, we evaluate four feature sets on the user conversion prediction task (Section~\ref{sec:experimental-setup}), all using the same MLP with binary cross-entropy loss, per-class positive weighting, and early stopping, so differences are attributable solely to input features:
(A) \textbf{Site retargeting features (SRF)}, 9 hand-crafted page-category counts over rolling windows;
(B) \textbf{distilled intents (student)}, the 35-dimensional probability vector from the distilled classifier (Section~\ref{sec:taxonomy});
(C) \textbf{LLM intents (teacher)}, the 35-dimensional raw confidence vectors from label assignment (Section~\ref{sec:taxonomy}); and
(D) \textbf{session embedding}, the 64-dimensional embedding (Section~\ref{sec:session-embedding}).

\begin{table}[t]
  \centering
  \caption{Conversion prediction results (micro metrics). To preserve business confidentiality, absolute values are withheld and percentage changes are reported relative to the LLM teacher baseline (C). Cell color encodes the change relative to the teacher (blue = improvement, red = degradation; darker is larger in magnitude, with the diverging scale saturating at a 20\% change).}
  \label{tab:conversion-results}
  \small
  \begin{tabular}{ll c r r r}
    \toprule
    & Model & Dim & Prec. & Rec. & F1 \\
    \midrule
    \multicolumn{6}{l}{\textit{Hand-crafted features}} \\
    A & Site Retargeting Features                  & 9   & \cellcolor{badred!60}$-$41.3\% & \cellcolor{badred!60}$-$23.6\% & \cellcolor{badred!60}$-$35.8\% \\
    \midrule
    \multicolumn{6}{l}{\textit{Intent-based representations}} \\
    B & Distilled (student)  & 35  & \cellcolor{badred!27}$-$8.9\%  & \cellcolor{badred!11}$-$3.6\%  & \cellcolor{badred!21}$-$7.0\% \\
    C & LLM intent (teacher) & 35  & --- & --- & --- \\
    \midrule
    \multicolumn{6}{l}{\textit{Self-supervised embedding}} \\
    D & \textbf{Session emb.}& \textbf{64} & \cellcolor{goodblue!22}\textbf{+7.3\%} & \cellcolor{badred!8}$-$0.5\% & \cellcolor{goodblue!13}\textbf{+4.3\%} \\
    \bottomrule
  \end{tabular}
\end{table}

\textbf{Learned representations outperform hand-crafted features.}
As shown in Table~\ref{tab:conversion-results}, SRF (A) shows a 35.8\% micro-F1 degradation relative to the LLM teacher (C), confirming that coarse page-category counts miss the behavioral nuance needed for conversion prediction.
The LLM teacher (C), our reference baseline for intent-based representations, encodes meaningful signal in its sparse 35-dimensional confidence vector; combining distilled intents with SRF (A+B) yields a +37\% relative F1 improvement over SRF alone.

\textbf{Distillation loses little; the raw embedding is richest.}
The distilled student (B) retains 93\% of the teacher's predictive power (a 7.0\% relative micro-F1 decrease) while operating on the 64-dimensional embedding, a tradeoff needed to scale from thousands of LLM calls to millions of MLP inferences (Section~\ref{sec:inference-efficiency}).
The session embedding (D) achieves the highest micro F1, a 4.3\% improvement over the teacher despite only 64 dimensions: it is the \emph{input} to the distillation pipeline, which extracts interpretable labels from the same space at some cost to information.
These results validate the dual-output design: the session embedding serves quantitative tasks with strong predictive power, while the distilled labels, from the same embedding space, add interpretability, segmentation, and stakeholder-facing explanations at little cost to accuracy.

\subsection{Inference Efficiency}
\label{sec:inference-efficiency}

A key motivation for distillation is production scalability.
Assigning intent labels via direct LLM inference (API or self-hosted) generates a full response per session at 2--4 seconds each; at 100K sessions this exceeds 80 hours of serial inference, infeasible for deployment.
By contrast, our pipeline is fast: the Transformer encodes 100K sessions into embeddings in 60 seconds, and the distilled MLP produces labels from them in 0.009 seconds, under 61 seconds per 100K sessions. This is a throughput improvement of more than 3,000$\times$ with only a 7\% drop in labeling performance, cutting the cost of large-scale batch processing while retaining a production-quality feature without separate text encoding or LLM inference.

\section{Conclusion and Future Work}

We presented a scalable system that turns web clickstream data into two outputs: session intent embeddings and interpretable intent labels. Together they link raw behavioral signals to a semantic reading of user intent in financial services. 
The session embeddings outperform baselines on the home screen tile ranking task, capturing richer signals than aggregation-based features, and the intent taxonomy provides actionable categories for segmentation, personalization, and marketing.
The distilled classifier also outperforms hand-crafted features while producing human-readable labels and enabling scalable inference, at a small and expected cost relative to the raw embeddings. The two models are a lightweight and deployable alternative to running an LLM on every session. They keep most of its intent understanding while using far less compute, and stay interpretable enough for regulated financial environments.

\textbf{Limitations.} The system is built and evaluated on clickstream data from a single financial institution; generalizability to other domains and tasks remains to be validated.

\textbf{Future work.}
We plan to (1) run online A/B tests to measure the impact of session embeddings and intent labels on tile ranking; (2) automate taxonomy refinement so the intent catalog adapts as behavior and products evolve; and (3) extend from session-level to user-level intent prediction by aggregating embeddings and labels across sessions per user over longer periods.

\bibliographystyle{plainnat}
\bibliography{references}

\end{document}